\def\40K{$^{40}$K}
\def\Li{$^{6}$Li}
\def\K{$^{39}$K}
\def\Na{$^{23}$Na}
\def\Rb{$^{87}$Rb}
\def\NaK{\Na\K}

\def\ket#1{\mathinner{|{#1}\rangle}}

\documentclass[aps,pra,twocolumn,
floatfix,10pt,superscriptaddress]{revtex4-2} 
\usepackage{graphicx,amsmath,amssymb,ulem}
\usepackage{textcomp}
\usepackage{hyperref}
\usepackage{color}
\usepackage{epstopdf}
\usepackage{nicefrac}
\usepackage[caption=false]{subfig}
\usepackage{gensymb}
\usepackage{placeins}
\usepackage{xspace}
\usepackage{lineno}

\begin{document}

\title{Hyperfine dependent atom-molecule loss analyzed by the analytic solution of few-body loss equations}

\author{Kai~K.~Voges}
\thanks{These authors contributed equally to this work.\\voges@iqo.uni-hannover.de}
\affiliation{Institut f\"ur Quantenoptik, Leibniz Universit\"at Hannover, 30167~Hannover, Germany}
\author{Philipp~Gersema}
\thanks{These authors contributed equally to this work.\\voges@iqo.uni-hannover.de}
\affiliation{Institut f\"ur Quantenoptik, Leibniz Universit\"at Hannover, 30167~Hannover, Germany}
\author{Torsten~Hartmann}
\affiliation{Institut f\"ur Quantenoptik, Leibniz Universit\"at Hannover, 30167~Hannover, Germany}
\author{Silke~Ospelkaus}
\email{silke.ospelkaus@iqo.uni-hannover.de}
\affiliation{Institut f\"ur Quantenoptik, Leibniz Universit\"at Hannover, 30167~Hannover, Germany}
\author{Alessandro~Zenesini}
\affiliation{Institut f\"ur Quantenoptik, Leibniz Universit\"at Hannover, 30167~Hannover, Germany}
\affiliation{INO-CNR BEC Center, Dipartimento di Fisica, Universit\`{a} di Trento\\ and TIFPA-INFN, 38123 Povo, Italy}

\date{\today}

\begin{abstract}
We prepare mixtures of ultracold \K\ atoms in various hyperfine spin states with \NaK\ molecules in an optical dipole trap at a fixed magnetic field  and study  inelastic two-body atom-molecule collisions. We observe atom-molecule losses that are hyperfine dependent with a two-body loss rate far below the universal limit.  We analyze the two-body loss dynamics based on the derivation of general and easy applicable analytic solutions for the differential equations describing the loss of an arbitrary number $\gamma$ of particles in a single collisional event. 

\end{abstract}
\maketitle
\section{Introduction}
Quantum gas experiments rely on the scattering properties of the neutral particles \cite{RevModPhys.71.1}. Their collisions are crucial for example in rethermalization during evaporative cooling \cite{PhysRevB.34.3476} or Bose-Einstein condensation \cite{Anderson198}. The control of the collisions by means of Feshbach resonances \cite{RevModPhys.82.1225} can lead to the selective emergence or collapse of Bose-Einstein condensates\cite{PhysRevLett.78.985,PhysRevLett.99.010403}, formation of liquid quantum droplets \cite{PhysRevLett.115.155302,Cabrera_2017} or creation of ultracold dimer molecules \cite{Donley_2002}. 
However, short-range collisions are often accompanied by losses. Especially resonant scattering conditions can enhance the losses from the ultracold ensemble. The loss dynamics itself offer valuable clues to the origin of these losses, for example the number of involved particles, the loss mechanisms or the density of resonant states \cite{PhysRevA.85.062712}.\\\indent
A prominent and timely example for few-body losses are ultracold heteronuclear ground-state molecular gases. In all so far created ground-state molecular many-particle gases two-body associated losses have been detected \cite{KRb1,GsDiMo87Rb133Cs2014Grimm,GsDiMo87Rb133Cs2014Cornish,GsDiMo23Na40K2015,NaRb12,GsDiMo23Na6Li2017,PhysRevLett.125.083401}, where the particular loss rate coefficients are close to the universal scattering limit \cite{C1CP21270B,Ospelkaus2010}. 
In the case of chemically unstable molecules, such as the KRb molecule, those losses originate mainly from two-body exothermic exchange reactions \cite{Ospelkaus2010,PhysRevA.81.060703,hu2019direct}. 
In contrast to that, chemically stable ground-state molecules, such as NaK, NaRb and RbCs are not affected by exothermic reactions \cite{PhysRevA.81.060703}, but still suffer from nearly universal loss in collisions \cite{GsDiMo23Na40K2015,NaRb12,GsDiMo87Rb133Cs2014Grimm,GsDiMo87Rb133Cs2014Cornish,PhysRevLett.125.083401}. This is suspected to be due to sticky collisions \cite{PhysRevA.87.012709,Gregory_2019}. Due to the high density of states in molecular collisions long-lived tetramer complexes form and might get excited by light from the optical dipole trap \cite{PhysRevLett.123.123402} such that both molecules get lost. This has been detected in the relatively heavy KRb and RbCs molecules \cite{liu2020steering, gregory2020loss}, but remains elusive in the light weight molecular systems of NaK and NaRb. Indeed, experiments with these molecules suggest complex lifetime several orders of magnitude larger than calculated \cite{gersema2021probing} or even additional unexplained loss mechanisms \cite{bause2021collisions}. 
The same kind of investigation recently expands to ensembles of molecule-atom systems, in which also sticky trimer complexes can form in collisions of a single dimer molecule with a single atom. Near universal losses due to light excitation have already been observed in KRb+Rb collisions \cite{nichols2021detection}, likewise for an unexpected high complex lifetime. Again, in contrast to the heavy systems, the light weight systems LiNa+Na and NaK+K have observed deviations from that behaviour, where losses can be suppressed \cite{son2019observation,PhysRevLett.125.083401} or even be tuned by means of external magnetic fields \cite{NaKYang261,wang2021magnetic}. The complex formation and the loss mechanisms are not fully understood and require further fundamental investigations towards few- and many-body quantum physics and chemistry. 
The data analysis of the loss dynamics involves the relatively complicated use of sets of differential equations modeling the loss of the particles and the associated temperature evolution in the trap. Numerical solving and fitting routines make a fast analysis of the data inconvenient and less intuitive.\\\indent
In this paper, we report the observation of hyperfine dependent atom-molecule loss in mixtures of  \K\ atoms prepared in various hyperfine spin states of the electronic ground state and bosonic \NaK\ molecules in a single but fixed hyperfine state of the rovibronic ground state. We observe the atom-molecule two-body loss rate to be dependent on the hyperfine spin state of the \K\ atoms and far below the universal limit. We analyze the two-body loss dynamics based on the derivation of analytic solutions to the differential equation system for the time evolution of particle number and temperature in a $\gamma$-body loss process.

In Sec. \ref{Sec:Model} we describe the general coupled differential equation  system for loss dynamics resulting from a $\gamma$-body loss process and the derivation of the analytic solutions to this problem. Sec. \ref{Sec:App} reports about our experimental work. First, we summarize the preparation of atom-molecule mixtures of \K\ atoms in various hyperfine spin states and \NaK\ molecules  in our experimental apparatus. We then discuss the hyperfine spin dependent loss measurements and their analysis using the analytic solution from Sec. \ref{Sec:Model} for the two-body decay ($\gamma=2$). Finally, we discuss our results in  Sec. \ref{Sec:Disc}.

\section{Analytical model for the $\gamma$-body loss problem}\label{Sec:Model}
The loss dynamics in pure molecular ensembles and atom-molecule mixtures presented in this paper follow typical two-body losses. This is described by a set of nonlinear differential equations where the loss rate coefficient $k_2$ is used as a fit parameter. 
Nevertheless, such a set of equations can be generally applied to the case of a $\gamma$-body problem, where $\gamma$ is the number of particles interacting with each other during a single scattering event. This is a universal problem in many fields of physics.

The typical approach to apply these systems to experimental data is to laboriously fit numerically deduced solutions of the differential equation system. The application of analytic solutions thus will be much more convenient and facilitate the analysis of experimental data drastically. In the following we derive these solutions for the loss problem in general with $\gamma$ involved particles. This makes our solutions applicable for a broad range of applications. \\

We consider a system of $N(t)$ particles of temperature $T(t)$ trapped in a harmonic potential $U(x,y,z)$ with trapping frequency $\omega_i$ for the direction $i$. Following the Boltzmann statistics, the size of the ensemble is given by a Gaussian profile with width $\sigma_i(t)=\sqrt{ k_B T(t)/m \omega_i^2)}$ \footnote{Note, we define $2\sigma$ as the distance, where the density drops to $\frac{1}{e^2}\cdot n(0,0,0,0)$}, where $ m$ is the mass of the particle and $k_B$ is the Boltzmann constant. The density of the ensemble is given by 
\begin{equation}\label{density}
 n(t,x,y,z)= N(t)\prod_{i=x,y,z}\frac{e^{-i^2/(2\sigma_i(t)^2)}}{\sqrt{2\pi}\sigma_i(t)}.
\end{equation}

If we limit the derivation to one- and $\gamma$-body particle loss processes in the system, with loss rates $k_1$ and $k_{\gamma}$, the particle number $N(t)$ and the temperature $T(t)$, assuming re-thermalization, result from a system of coupled equations
\begin{equation}
\begin{array}{*{20}{l}} {\dot N(t)} \hfill & = \hfill & { - k_{\gamma} \langle n(t)^\gamma\rangle _V-k_1 N(t)}\,,\\ {\dot T(t)} \hfill & = \hfill & { + \frac{\Gamma_\gamma(t)}{3k_B N(t)} k_{\gamma} \langle n(t)^\gamma\rangle _V} \,, \hfill \end{array}
\label{Diff}
\end{equation}
where the $\langle n(t)^\gamma \rangle_V$ is the volume integral of the $\gamma$-body density $n(t,x,y,z)^\gamma$. The anti-evaporation is described by the quantity $\Gamma_\gamma(t)$ and corresponds to the mean potential energy
\begin{equation}
\Gamma_\gamma(t) =\frac{3}{2} k_B T(t) - \langle U(x,y,z) n(t,x,y,z)^\gamma \rangle _V \,.
\end{equation}
$\Gamma_\gamma(t)$ depends on time both through the temperature and the density.
Note that the one-body loss process does not contribute to the anti-evaporation as it is not density-dependent.

By explicitly integrating the volume integrals over the Gaussian density profiles, one reaches the following system of equations
\begin{equation}
\begin{array}{*{20}{l}} {\dot N(t)} \hfill & = \hfill & { - \frac{k_{\gamma} C^{(\gamma - 1)}}{\gamma ^{3/2}}\ {\frac{{N(t)^\gamma }}{{T(t)^{(3/2)(\gamma - 1)}}}}-k_1 N(t)\,,} \hfill \\ {\dot T(t)} \hfill & = \hfill & {\frac{k_{\gamma} C^{(\gamma - 1)}}{\gamma ^{3/2}}\left( {\frac{{\gamma - 1}}{{2\gamma }}} \right) {\frac{{N(t)^{\gamma - 1}}}{{T(t)^{(3\gamma - 5)/2}}}}\,,} \hfill \end{array}
\label{Diff}
\end{equation}
where $C$ is $(m \omega^2/(2\pi k_B))^{(3/2)}$ and $\omega$ is the average trap frequency of the trapping potential $\omega=(\omega_x\omega_y\omega_z)^{1/3}$. Note that we use the same conventions as used in \cite{Gregory_2019}.

The first step to an analytic solution for $N(t)$ and $T(t)$ consists in substituting $N(t)$ with $M(t)e^{-k_1t}$ to hide the one-body term
\begin{equation}
\begin{array}{*{20}{l}} {\dot M(t)} \hfill & = \hfill & { - \frac{k_{\gamma} C^{(\gamma - 1)}}{\gamma ^{3/2}} {\frac{{M(t)^\gamma }}{{T(t)^{(3/2)(\gamma - 1)}}}} e^{-(\gamma-1) k_1t},} \hfill \\ {\dot T(t)} \hfill & = \hfill & {\frac{k_{\gamma} C^{(\gamma - 1)}}{\gamma ^{3/2}} {\frac{{\gamma - 1}}{{2\gamma }}} \left( {\frac{{M(t)^{\gamma - 1}}}{{T(t)^{(3\gamma - 5)/2}}}} \right)e^{-(\gamma-1)k_1t}.} \hfill 
\end{array}
\label{Diff}
\end{equation}
By dividing the first line by the second one in Eq.\,\ref{Diff} we have:
\begin{equation}
\frac{\dot M(t)}{\dot T(t)}= -\frac{2\gamma}{\gamma-1}\frac{M(t)}{T(t)},
\end{equation}
which leads to 
\begin{equation}
\frac{M(t)}{N_0}= \left( \frac{T(t)}{T_0 }\right)^{-\frac{2\gamma}{\gamma-1}},
\label{NoverT}
\end{equation}
where $N_0=N(0)=M(0)$ and $T_0=T(0)$ are the initial atom number and the initial temperature.
By introducing $M(t)$ and $N_0$ in the first equation in Eq.\,\ref{Diff}, we get

\begin{equation}
\frac{{\dot M(t)}}{N_0} \hfill  =\hfill  { -  N_0^{\gamma-1} \frac{k_{\gamma} C^{(\gamma - 1)}}{\gamma ^{3/2}}{\frac{{(M(t)/N_0)^\gamma }}{{T(t)^{(3/2)(\gamma - 1)}}}} e^{-(\gamma-1) k_1t},}
\end{equation}
where $M(t)/N_0$ can be substituted using Eq.\,\ref{NoverT}:

\begin{align}
&\frac{d}{dt}\left[ \left( \frac{T(t)}{T_0 }\right)^{-\frac{2\gamma}{\gamma-1}}\right] \hfill  =\\ &\hfill   { - \frac{k_{\gamma} (N_0C)^{\gamma - 1}}{\gamma ^{3/2}} {\frac{{\left(\left( \frac{T(t)}{T_0 }\right)^{-\frac{2\gamma}{\gamma-1}}\right)^\gamma }}{{T(t)^{(3/2)(\gamma - 1)}}}} e^{-(\gamma-1)k_1t}.} \nonumber
\end{align}

This differential equation for $T(t)$ has the solution
\begin{equation}
\frac{T(t)}{T_0} = \left( 1+ k_\gamma\frac{\eta}{\alpha} \left(\frac{1-e^{-(\gamma-1)k_1t}}{(\gamma-1)k_1}\right)\right)^\beta,
\label{SolT}
\end{equation}
where $\eta$ contains the starting conditions and trapping parameters
\begin{equation}
\eta=\frac{1}{\gamma^{3/2}}\left(\frac{C N_0}{T_0^{3/2}}\right)^{\gamma-1}.
\end{equation}
The two constants $\alpha$ and $\beta$ depend only on $\gamma$ and can be written as  
\begin{equation}
\alpha=\frac{4\gamma}{(\gamma-1)(7\gamma-3)}\,,
\end{equation}
\begin{equation}
\beta=\frac{2}{7\gamma-3}\,.
\end{equation}
The solution for $N(t)$ is then obtained by combining Eq.\,\ref{NoverT} and Eq.\,\ref{SolT}.

\begin{equation}
\frac{N(t)}{N_0} = \frac{e^{-k_1t}}{\left( 1+k_\gamma \frac{\eta}{\alpha} \left(\frac{1-e^{-(\gamma-1)k_1t}}{(\gamma-1)k_1}\right)\right)^{\alpha}}\, .
\label{SolN}
\end{equation}

For $\gamma=1$, this solution collapses to the known exponential decay. The solutions for $\gamma=2,3,4,5$ have the following $\alpha$ and $\beta$:

\begin{center}
\begin{tabular}{ c | c  | c }
$\gamma$ & $\alpha$ & $\beta$\\
\hline
2 & 8/11 & 2/11 \\
3 & 1/3 & 1/9 \\
4 & 16/75 & 2/25 \\
5 & 5/32 & 1/16 \\

\end{tabular}
\end{center}
Note that the case $\gamma=3$ corresponds to the solution already presented by Kraemer \cite{KraemerPhillip-Tobias2006Fiia} although the solution of $T(t)$ in the thesis manuscript has a typo on $\eta$.
The limit for negligible one-body losses ($k_1\to 0$) can be obtained just by noticing that  $\lim_{k_1 \to 0}\left(\frac{1-e^{-(\gamma-1)k_1t}}{(\gamma-1)k_1}\right)=t$, which leads to:
\begin{equation}
\begin{array}{*{20}{l}}
T(t)/T_0 & = \left( 1+ k_\gamma \frac{\eta}{\alpha} t\right)^\beta, \\
N(t)/N_0 & = \left( 1+k_\gamma \frac{\eta}{\alpha}t \right)^{-\alpha}.
\label{SolNT}
\end{array}
\end{equation}
In the following,  we use our newly derived analytic solutions to analyze the hyperfine dependent atom-molecule loss observed in our experiments.

\section{Hyperfine  dependent atom-molecule loss }\label{Sec:App}
\subsection{Preparation of atom-molecule mixtures}\label{SubSec:Prep}
The ground-state molecule creation starts from mixtures of about $1.8\times 10^5$ ultracold \Na\ and $0.6\times 10^5$ \K\ atoms trapped in a 1064\,nm crossed beam optical dipole trap (cODT) with temperatures of 300\,nK. The cODT frequencies are measured by trap oscillations of the respective particles. For \K\ atoms the trap frequencies are $\omega_{\textrm{K},(x,y,z)}=2\pi\times(375.8(8.6), 375.8(8.6), 60.8(6.4))\,$Hz. The trap frequencies for \Na\ atoms are scaled by a factor of about $0.62$ due to different dynamical polarizability and mass. The atoms are initially prepared in the states $\ket{f=1, m_f=-1}_\textrm{Na}+\ket{f=1, m_f=-1}_\textrm{K}$ for which the inter- and intra-species scattering properties are well known \cite{SchulzeBEC2018}. Feshbach molecules are formed close to a Feshbach resonance at about 200\,G in the atomic $\ket{f=1, m_f=-1}_\textrm{Na}+\ket{f=2, m_f=-2}_\textrm{K}$ states \cite{Voges2019Fesh}. The bound molecular state is populated by a short radio frequency pulse of 400\,$\mu$s. Immediately after molecule formation the transfer to the ground state takes place. This is done within 12\,$\mu$s via a stimulated Raman adiabatic passage (STIRAP) to a single rovibrational hyperfine state $\ket{m_{i,\text{Na}}=-3/2,m_{i,\text{K}}=-1/2,M_J=0,M_{i,\text{NaK}}=-2}$\,\,\cite{PhysRevLett.125.083401}. For detection of molecules the STIRAP is reversed and \K\ atoms from the Feshbach molecule state are imaged via a cycling transition. Typically, about 4000 ground-state molecules are generated per experimental cycle. The determined trap frequencies of the \NaK\ molecules are $\omega_{\textrm{NaK},(x,y,z)}=2\pi\times(195.6(8.3), 195.6(8.3), 33.1(1.3))\,$Hz. The temperature of the \NaK\ molecules is measured by a time-of-flight measurement on the \K\ atoms after STIRAP reversal and Feshbach molecule dissociation \cite{PhysRevLett.125.083401}. The temperature of the remaining \K\ atoms is still 300\,nK, from which we conclude that neither molecule association nor dissociation lead to heating of the molecules.
After ground-state transfer the molecules are still embedded in a bath of remaining atoms. \Na\ atoms are usually removed as fast as possible since the molecules undergo a chemical reaction with \Na\ atoms. This is done by a 500\,$\mu$s resonant light pulse, which pushes the atoms out of the cODT. To prevent optical pumping and thus incomplete removal of the \Na\ atoms, two resonant frequencies are used connecting the excited state $\ket{f=2,m_f=-2}_\textrm{Na}$ to the corresponding low field ground states  $\ket{1,-1}_\textrm{Na}$ and $\ket{2,-1}_\textrm{Na}$. The notation is given in the basis of the Zeeman regime with $\ket{f,m_f}$, where $f$ is the hyperfine number and $m_f$ its projection.
After the removal of the \Na\ atoms, the molecules are left in a gas of \K\ atoms in the  $\ket{1,-1}$ hyperfine state. Fortunately, this atomic state leads to very little losses of the molecules \cite{PhysRevLett.125.083401}, which is favorable for all further state manipulations of the mixture; see Fig.\,\ref{MoleAtomLoss}. For the preparation of a pure molecular cloud \K\ atoms are removed from the trap by a rapid adiabatic passage (RAP) to the $\ket{2,-2}$ state followed by a $500\,\mu\textrm{s}$ resonant light pulse. For the study of atom-molecule loss with \K\ atoms in various spin states, the hyperfine state of the \K\ atoms is manipulated using  RAPs and rf pulses. In our experiment, we  prepare six different \K\ hyperfine states: $\ket{1,1}$, $\ket{1,0}$, $\ket{1,-1}$, $\ket{2,-2}$, $\ket{2,-1}$ and $\ket{2,0}$, where the states are written in the basis $\ket{f,m_f}$.
\subsection{Atom-molecule two-body loss coefficients}\label{SubSec:Ana}
Atom-molecule collisions between \NaK\ molecules in the $\ket{m_{i,\text{Na}}=-3/2,m_{i,\text{K}}=-1/2,M_J=0,M_{i,\text{NaK}}=-2}$ state \cite{PhysRevLett.125.083401} and \K\ atoms in one of the six different hyperfine states mentioned above are studied in the optical dipole trap at a fixed magnetic field of 200\,G in the vicinity of the used Feshbach resonance.  Apart from losses due to two-body atom-molecule collisions, the decay dynamics of  an atom-molecule mixture is also affected by the two-body losses due to molecule-molecule collisions, and it is therefore important to first understand in detail the two-body decay of a pure molecular cloud. 

\begin{figure}[t]
	\includegraphics[width=1\columnwidth]{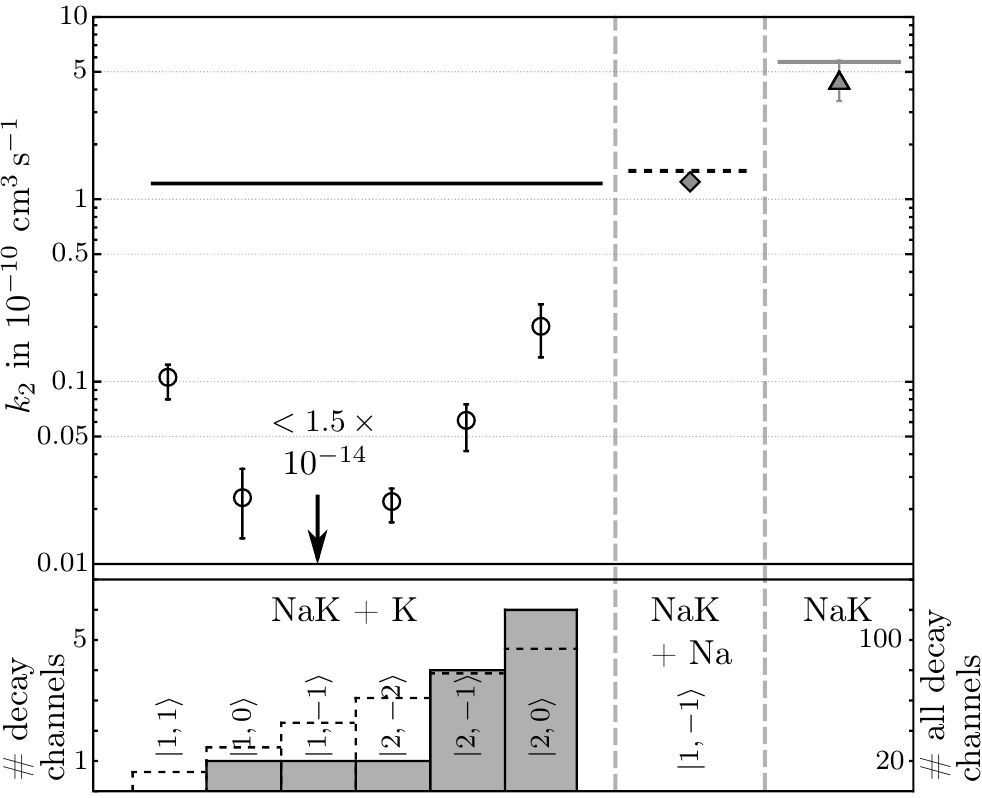}
	\caption{Loss in atom-molecule mixtures. The upper part shows the  two-body loss rate coefficients for the different molecular systems. The left panel includes the loss rate coefficients for the \NaK+\K\ mixture with different hyperfine states for \K\ atoms (open circles). The values for \NaK+\Na\ (gray diamond, middle panel) and \NaK\ (gray triangle, right panel) are shown for comparison. Note, that the value for the $\ket{1,-1}$ state is so small, that it is not shown in the graph. All measurements were performed at a temperature of 300\,nK. The horizontal lines represent the respective universal limit, calculated also for 300\,nK. The solid gray bar diagram in the lower part shows the number of available loss channels (left axes) due to hyperfine changing collisions for the \NaK+\K\ mixture with the total $M_F=M_{i,\text{NaK}}+m_{f,{\mathrm{K}}}$ preserved. The dashed bar diagram indicates the number of all energetically possible decay channels (right axes), if one assumes that $M_F$ is not preserved.}
	\label{MoleAtomLoss}
\end{figure}

Pure molecular collisions between \NaK\ molecules have already been studied in our previous work \cite{PhysRevLett.125.083401}. One-body background losses for example from vacuum background gas collisions \cite{SchulzeBEC2018} or photo-excitation of bialkali molecules are negligible \cite{PhysRevA.78.052901} and the dominating loss is observed to result from two-body molecule-molecule collisions. We thus analyse the decay of a pure molecular cloud  using the analytic solution from Eq.\,\ref{SolNT}  with  $\gamma=2$. Accordingly, $\alpha$ is $8/11$ and the constant $\eta$ contains the parameters of the particles and the trap, which are the initial particle number and temperature as well as trap frequencies; see Sec.\,\ref{SubSec:Prep}. We fit this model  to the decay of a pure molecular cloud and extract the two-body loss rate coefficient $k_2=4.49(1.18)\times 10^{-10} \,\textrm{cm}^3 \textrm{s}^{-1}$ at 200\,G close to the universal limit and consistent with the value reported in \cite{PhysRevLett.125.083401} (see Fig. \ref{MoleAtomLoss}).\\

Knowing the background two-body loss resulting from collisions of two ground-state molecules we can investigate ultracold collisions of molecules with \K\ atoms in various hyperfine states.
We therefore prepare different atom-molecule mixtures of \K\ atoms in one of the six different hyperfine states and \NaK\  molecules; see Sec. \ref{SubSec:Prep}. We measure the decay of the molecular cloud as a function of time for each of the six different atom-molecule mixtures and analyze loss dynamics  using solutions from Sec.\,\ref{Sec:Model}, in particular Eq.\,\ref{SolN}. 

To apply Eq.\,\ref{SolN} to this problem, one needs to include potassium losses into the analytical model.
Given that the density of the \K\ gas is one to two orders of magnitude larger than the density of the molecular gas, we can safely assume that the atomic density 
does not change during the decay of the molecular cloud. Therefore one can treat this like losses from vacuum background collisions, although that does not account for possible re-thermalization of the molecules with the potassium cloud. This results in a molecular two-body decay with an additional exponential one-body decay on top, originating from a two-body atom-molecule loss.

We thus analyze the loss dynamics of the molecular cloud 
considering molecule-molecule collisions with the obtained $k_2$ and modeling atom-molecule loss as a  one-body (molecule) loss part with  $k_1$ from Eq.\,\ref{SolN}  expressed as
\begin{equation}
k_1=k_{2,\textrm{K}}n_\textrm{K}\, ,
\end{equation}
where $n_\textrm{K}$ is the initial \K\ density and $k_{2,\textrm{K}}$ the actual two-body loss rate coefficient, including spatial density overlap of the atomic and molecular cloud in the ODT, for the atom-molecule collision. \\
Note that, within the experimental and statistical uncertainties, the obtained values of $k_1$, found with the solutions in Sec.\,II, are compatible with the ones calculated for a system where the atoms and molecules thermalize and therefore remain at a fixed temperature.\\
We extract the $k_{2,\textrm{K}}$ loss rate coefficients for the atom-molecule collision for the six different hyperfine states of the \K\ atoms.  The results are shown in Fig.\,\ref{MoleAtomLoss} in the left panel.  All state combinations show a loss rate coefficient much smaller than the calculated universal limit \cite{PhysRevA.101.063605}, with one state even showing a drastic suppression of the loss rate coefficient of several orders of magnitude \cite{PhysRevLett.125.083401}. For comparison, we include the calculated universal scattering limit as a solid black line in Fig.\,\ref{MoleAtomLoss},  the two-body loss rate coefficient for \NaK+\Na\ (Fig.\,\ref{MoleAtomLoss}, middle panel) as well as the molecule-molecule two-body loss rate coefficient (right panel). The lower part of Fig. 1 shows the number of available energetically allowed loss channels due to hyperfine changing collisions with and without preserved $M_F$ (left and right axes, respectively).
The loss rate coefficient for \NaK+\Na\ collisions obtained by mean of our model  is in agreement with the universal limit and with the fact that for a chemical unstable reaction, all collisions lead to loss of the involved atom and molecule. This is consistent with the large losses observed during the sample preparation and the aforementioned necessity to remove \Na\ as quickly as possible from the trap. The numerical agreement also allows us to verify that the applied model gives reasonable output.

\section{Discussion}\label{Sec:Disc}
Without including hyperfine interactions and possible coupling to trimer states, the endoergic nature of the \NaK+\K\ system \cite{PhysRevA.81.060703} would result in stable and long-lived mixtures, as there are no two-body losses from chemical exchange reactions.\\\indent
However, near-universal decay which has been attributed to the coupling to a large density of trimer states has been observed in various atom-molecule systems with the exception of \40K\Rb+\Rb\ mixtures when both atoms and molecules  have been prepared in their lowest stretched states \cite{Ospelkaus2010} and in \Li\Na+\Na\ with both particles in their highest stretched states \cite{son2019observation}. In other state combinations of both examples losses consistent with the universal limit have been reported \cite{Ospelkaus2010,GsDiMo23Na6Li2017}.
Meanwhile, for \Na\40K+\40K\ collisions numerous magnetic field tuneable Feshbach resonances in different hyperfine states have been found and assigned to long-range states \cite{wang2021magnetic}. In fact, in these experiments collisional resonances have been identified as an enhancement of the loss rate coefficient in the vicinity of a resonance starting from a near-universal background loss \cite{NaKYang261}.

In contrast, for the bosonic \NaK+\K\ case we do report suppression of losses far below the universal limit where neither the molecules nor the atoms are in their stretched states at the same time, see Fig.\,\ref{MoleAtomLoss}. Although for the bosonic case \NaK+\K\ no magnetic field dependent measurements have been done so far, still the change of the loss rate coefficient for different spin channels suggests a hyperfine dependent Feshbach structure, which is similarly tuneable as in fermionic \Na\40K+\40K\ collisions.\\\indent
The effect of the changed loss rate coefficients and the loss of the particles remains elusive. We rule out, that the losses of molecules originate from hyperfine changing collisions between atoms and molecules with preserved total $M_F$ of the atom-molecule system.  For this purpose we determine the Zeeman structure of the lowest hyperfine states of the  \NaK+\K\ at the magnetic field of 200\,G, including hyperfine structure of the molecule in the Paschen-Back and for the atom in the intermediate regime and compare the number of possible $M_F=M_{i,\text{NaK}}+m_{f,{\mathrm{K}}}$ preserving inelastic loss channels for atom-molecule mixture with the measured loss rate coefficients.  The results are shown in the lower panel in Fig.\,\ref{MoleAtomLoss}. The number of available hyperfine changing loss channels does not correlate with the loss rate coefficients. This indicates that the observed loss is not dominated by hyperfine changing collisional processes.

In dipolar collisions, hyperfine changing processes can also lead to a non-preserved total angular momentum projection $M_F$ \cite{sonding1998, Hensler2003}.  The dipolar angular momentum of the \NaK\ molecule in absence of an externally applied electric field is zero. Moreover, the only dipole moments in our \NaK\ system are the nuclear ones, which are only relevant at short range. Since we cannot exclude the possibility that some resonant processes amplify the influence of the nuclear dipoles, we give in Fig.\,\ref{MoleAtomLoss} the number of all energetically possible decay channels even when $M_F$ is not preserved. We find that this number does not correlate with the loss rate coefficients, thus we can rule out a role of hyperfine changing processes just based on possible decay channels.

One other possibility of losses from a collisional system are light-induced losses of long-lived trimer complexes \cite{PhysRevLett.123.123402, nichols2021detection}, for example from the light of the optical trap. Estimations of rovibrational as well as the hyperfine spacing for the \NaK+\K\ complex give evidence for resolvable photo-excitation resonances which might explain the increased loss rate coefficient in certain hyperfine states of \K\ atoms \cite{Hutsonprivate}. 

\section{Conclusion}\label{Sec:Con}
In summary we investigated atom-molecule two-body loss in a mixture of ultracold  \NaK\ molecules and \K\ atoms in six different hyperfine states prepared in an optical dipole trap at a fixed magnetic field of around 200 G.  We  observed  atom-molecule loss that is hyperfine dependent with a two-body loss rate far below the universal limit.  We analyzed the two-body loss dynamics based on the derivation of general and easily applicable analytic solutions for the differential equations describing the loss of $\gamma$ particles  in a single collisional event, which are easily implementable in data analysis. In contrast to the numerous atom-molecule mixtures studied to date, in which losses  predominantly near or at the universal limit have been observed, the light NaK+K system exhibits interesting properties that require further investigation. Feshbach resonances and hyperfine-dependent collisions have been observed and are far from being understood. New theory insight will be required to understand the physical mechanisms underlying the observed phenomena. Such insight will help to develop a complete understanding of atom-molecule collisions and associated trimer formation and should ultimately be applicable also to molecule-molecule collisions.  Detailed understanding of both molecule-molecule and atom-molecule collisions is crucial for further advancement in the production of molecular gases close to quantum degeneracy and associated research opportunities in quantum chemistry, quantum simulation, quantum information and precision measurements.
\newline
\section*{Acknowledgements}
We thank Goulven Qu{\'e}m{\'e}ner for enlightening comments and Jeremy Hutson and Matthew Frye for insightful conceptions. We gratefully acknowledge financial support from the European Research Council through ERC Starting Grant POLAR and from the Deutsche Forschungsgemeinschaft (DFG) through CRC 1227 (DQ-mat), project A03 and FOR2247, project E5. P.G. thanks the Deutsche Forschungsgemeinschaft for financial support through Research Training Group 1991. A.Z. thanks Provincia Autonoma di Trento for financial support.

\bibliography{Collisions}

\begin{thebibliography}{41}%
\makeatletter
\providecommand \@ifxundefined [1]{%
 \@ifx{#1\undefined}
}%
\providecommand \@ifnum [1]{%
 \ifnum #1\expandafter \@firstoftwo
 \else \expandafter \@secondoftwo
 \fi
}%
\providecommand \@ifx [1]{%
 \ifx #1\expandafter \@firstoftwo
 \else \expandafter \@secondoftwo
 \fi
}%
\providecommand \natexlab [1]{#1}%
\providecommand \enquote  [1]{``#1''}%
\providecommand \bibnamefont  [1]{#1}%
\providecommand \bibfnamefont [1]{#1}%
\providecommand \citenamefont [1]{#1}%
\providecommand \href@noop [0]{\@secondoftwo}%
\providecommand \href [0]{\begingroup \@sanitize@url \@href}%
\providecommand \@href[1]{\@@startlink{#1}\@@href}%
\providecommand \@@href[1]{\endgroup#1\@@endlink}%
\providecommand \@sanitize@url [0]{\catcode `\\12\catcode `\$12\catcode
  `\&12\catcode `\#12\catcode `\^12\catcode `\_12\catcode `\%12\relax}%
\providecommand \@@startlink[1]{}%
\providecommand \@@endlink[0]{}%
\providecommand \url  [0]{\begingroup\@sanitize@url \@url }%
\providecommand \@url [1]{\endgroup\@href {#1}{\urlprefix }}%
\providecommand \urlprefix  [0]{URL }%
\providecommand \Eprint [0]{\href }%
\providecommand \doibase [0]{https://doi.org/}%
\providecommand \selectlanguage [0]{\@gobble}%
\providecommand \bibinfo  [0]{\@secondoftwo}%
\providecommand \bibfield  [0]{\@secondoftwo}%
\providecommand \translation [1]{[#1]}%
\providecommand \BibitemOpen [0]{}%
\providecommand \bibitemStop [0]{}%
\providecommand \bibitemNoStop [0]{.\EOS\space}%
\providecommand \EOS [0]{\spacefactor3000\relax}%
\providecommand \BibitemShut  [1]{\csname bibitem#1\endcsname}%
\let\auto@bib@innerbib\@empty
\bibitem [{\citenamefont {Weiner}\ \emph {et~al.}(1999)\citenamefont {Weiner},
  \citenamefont {Bagnato}, \citenamefont {Zilio},\ and\ \citenamefont
  {Julienne}}]{RevModPhys.71.1}%
  \BibitemOpen
  \bibfield  {author} {\bibinfo {author} {\bibfnamefont {J.}~\bibnamefont
  {Weiner}}, \bibinfo {author} {\bibfnamefont {V.~S.}\ \bibnamefont {Bagnato}},
  \bibinfo {author} {\bibfnamefont {S.}~\bibnamefont {Zilio}},\ and\ \bibinfo
  {author} {\bibfnamefont {P.~S.}\ \bibnamefont {Julienne}},\ }\bibfield
  {title} {\bibinfo {title} {Experiments and theory in cold and ultracold
  collisions},\ }\href {https://doi.org/10.1103/RevModPhys.71.1} {\bibfield
  {journal} {\bibinfo  {journal} {Rev. Mod. Phys.}\ }\textbf {\bibinfo {volume}
  {71}},\ \bibinfo {pages} {1} (\bibinfo {year} {1999})}\BibitemShut {NoStop}%
\bibitem [{\citenamefont {Hess}(1986)}]{PhysRevB.34.3476}%
  \BibitemOpen
  \bibfield  {author} {\bibinfo {author} {\bibfnamefont {H.~F.}\ \bibnamefont
  {Hess}},\ }\bibfield  {title} {\bibinfo {title} {Evaporative cooling of
  magnetically trapped and compressed spin-polarized hydrogen},\ }\href
  {https://doi.org/10.1103/PhysRevB.34.3476} {\bibfield  {journal} {\bibinfo
  {journal} {Phys. Rev. B}\ }\textbf {\bibinfo {volume} {34}},\ \bibinfo
  {pages} {3476} (\bibinfo {year} {1986})}\BibitemShut {NoStop}%
\bibitem [{\citenamefont {Anderson}\ \emph {et~al.}(1995)\citenamefont
  {Anderson}, \citenamefont {Ensher}, \citenamefont {Matthews}, \citenamefont
  {Wieman},\ and\ \citenamefont {Cornell}}]{Anderson198}%
  \BibitemOpen
  \bibfield  {author} {\bibinfo {author} {\bibfnamefont {M.~H.}\ \bibnamefont
  {Anderson}}, \bibinfo {author} {\bibfnamefont {J.~R.}\ \bibnamefont
  {Ensher}}, \bibinfo {author} {\bibfnamefont {M.~R.}\ \bibnamefont
  {Matthews}}, \bibinfo {author} {\bibfnamefont {C.~E.}\ \bibnamefont
  {Wieman}},\ and\ \bibinfo {author} {\bibfnamefont {E.~A.}\ \bibnamefont
  {Cornell}},\ }\bibfield  {title} {\bibinfo {title} {\textrm{Observation of
  Bose-Einstein Condensation in a Dilute Atomic Vapor}},\ }\href
  {https://doi.org/10.1126/science.269.5221.198} {\bibfield  {journal}
  {\bibinfo  {journal} {Science}\ }\textbf {\bibinfo {volume} {269}},\ \bibinfo
  {pages} {198} (\bibinfo {year} {1995})}\BibitemShut {NoStop}%
\bibitem [{\citenamefont {Chin}\ \emph {et~al.}(2010)\citenamefont {Chin},
  \citenamefont {Grimm}, \citenamefont {Julienne},\ and\ \citenamefont
  {Tiesinga}}]{RevModPhys.82.1225}%
  \BibitemOpen
  \bibfield  {author} {\bibinfo {author} {\bibfnamefont {C.}~\bibnamefont
  {Chin}}, \bibinfo {author} {\bibfnamefont {R.}~\bibnamefont {Grimm}},
  \bibinfo {author} {\bibfnamefont {P.}~\bibnamefont {Julienne}},\ and\
  \bibinfo {author} {\bibfnamefont {E.}~\bibnamefont {Tiesinga}},\ }\bibfield
  {title} {\bibinfo {title} {Feshbach resonances in ultracold gases},\ }\href
  {https://doi.org/10.1103/RevModPhys.82.1225} {\bibfield  {journal} {\bibinfo
  {journal} {Rev. Mod. Phys.}\ }\textbf {\bibinfo {volume} {82}},\ \bibinfo
  {pages} {1225} (\bibinfo {year} {2010})}\BibitemShut {NoStop}%
\bibitem [{\citenamefont {Bradley}\ \emph {et~al.}(1997)\citenamefont
  {Bradley}, \citenamefont {Sackett},\ and\ \citenamefont
  {Hulet}}]{PhysRevLett.78.985}%
  \BibitemOpen
  \bibfield  {author} {\bibinfo {author} {\bibfnamefont {C.~C.}\ \bibnamefont
  {Bradley}}, \bibinfo {author} {\bibfnamefont {C.~A.}\ \bibnamefont
  {Sackett}},\ and\ \bibinfo {author} {\bibfnamefont {R.~G.}\ \bibnamefont
  {Hulet}},\ }\bibfield  {title} {\bibinfo {title} {\textrm{Bose-Einstein
  Condensation of Lithium: Observation of Limited Condensate Number}},\ }\href
  {https://doi.org/10.1103/PhysRevLett.78.985} {\bibfield  {journal} {\bibinfo
  {journal} {Phys. Rev. Lett.}\ }\textbf {\bibinfo {volume} {78}},\ \bibinfo
  {pages} {985} (\bibinfo {year} {1997})}\BibitemShut {NoStop}%
\bibitem [{\citenamefont {Roati}\ \emph {et~al.}(2007)\citenamefont {Roati},
  \citenamefont {Zaccanti}, \citenamefont {D'Errico}, \citenamefont {Catani},
  \citenamefont {Modugno}, \citenamefont {Simoni}, \citenamefont {Inguscio},\
  and\ \citenamefont {Modugno}}]{PhysRevLett.99.010403}%
  \BibitemOpen
  \bibfield  {author} {\bibinfo {author} {\bibfnamefont {G.}~\bibnamefont
  {Roati}}, \bibinfo {author} {\bibfnamefont {M.}~\bibnamefont {Zaccanti}},
  \bibinfo {author} {\bibfnamefont {C.}~\bibnamefont {D'Errico}}, \bibinfo
  {author} {\bibfnamefont {J.}~\bibnamefont {Catani}}, \bibinfo {author}
  {\bibfnamefont {M.}~\bibnamefont {Modugno}}, \bibinfo {author} {\bibfnamefont
  {A.}~\bibnamefont {Simoni}}, \bibinfo {author} {\bibfnamefont
  {M.}~\bibnamefont {Inguscio}},\ and\ \bibinfo {author} {\bibfnamefont
  {G.}~\bibnamefont {Modugno}},\ }\bibfield  {title} {\bibinfo {title}
  {\textrm{$^{39}\mathrm{K}$ Bose-Einstein Condensate with Tunable
  Interactions}},\ }\href {https://doi.org/10.1103/PhysRevLett.99.010403}
  {\bibfield  {journal} {\bibinfo  {journal} {Phys. Rev. Lett.}\ }\textbf
  {\bibinfo {volume} {99}},\ \bibinfo {pages} {010403} (\bibinfo {year}
  {2007})}\BibitemShut {NoStop}%
\bibitem [{\citenamefont {Petrov}(2015)}]{PhysRevLett.115.155302}%
  \BibitemOpen
  \bibfield  {author} {\bibinfo {author} {\bibfnamefont {D.~S.}\ \bibnamefont
  {Petrov}},\ }\bibfield  {title} {\bibinfo {title} {\textrm{Quantum Mechanical
  Stabilization of a Collapsing Bose-Bose Mixture}},\ }\href
  {https://doi.org/10.1103/PhysRevLett.115.155302} {\bibfield  {journal}
  {\bibinfo  {journal} {Phys. Rev. Lett.}\ }\textbf {\bibinfo {volume} {115}},\
  \bibinfo {pages} {155302} (\bibinfo {year} {2015})}\BibitemShut {NoStop}%
\bibitem [{\citenamefont {Cabrera}\ \emph {et~al.}(2018)\citenamefont
  {Cabrera}, \citenamefont {Tanzi}, \citenamefont {Sanz}, \citenamefont
  {Naylor}, \citenamefont {Thomas}, \citenamefont {Cheiney},\ and\
  \citenamefont {Tarruell}}]{Cabrera_2017}%
  \BibitemOpen
  \bibfield  {author} {\bibinfo {author} {\bibfnamefont {C.~R.}\ \bibnamefont
  {Cabrera}}, \bibinfo {author} {\bibfnamefont {L.}~\bibnamefont {Tanzi}},
  \bibinfo {author} {\bibfnamefont {J.}~\bibnamefont {Sanz}}, \bibinfo {author}
  {\bibfnamefont {B.}~\bibnamefont {Naylor}}, \bibinfo {author} {\bibfnamefont
  {P.}~\bibnamefont {Thomas}}, \bibinfo {author} {\bibfnamefont
  {P.}~\bibnamefont {Cheiney}},\ and\ \bibinfo {author} {\bibfnamefont
  {L.}~\bibnamefont {Tarruell}},\ }\bibfield  {title} {\bibinfo {title}
  {\textrm{Quantum liquid droplets in a mixture of Bose-Einstein
  condensates}},\ }\href {https://doi.org/10.1126/science.aao5686} {\bibfield
  {journal} {\bibinfo  {journal} {Science}\ }\textbf {\bibinfo {volume}
  {359}},\ \bibinfo {pages} {301–304} (\bibinfo {year} {2018})}\BibitemShut
  {NoStop}%
\bibitem [{\citenamefont {Donley}\ \emph {et~al.}(2002)\citenamefont {Donley},
  \citenamefont {Claussen}, \citenamefont {Thompson},\ and\ \citenamefont
  {Wieman}}]{Donley_2002}%
  \BibitemOpen
  \bibfield  {author} {\bibinfo {author} {\bibfnamefont {E.~A.}\ \bibnamefont
  {Donley}}, \bibinfo {author} {\bibfnamefont {N.~R.}\ \bibnamefont
  {Claussen}}, \bibinfo {author} {\bibfnamefont {S.~T.}\ \bibnamefont
  {Thompson}},\ and\ \bibinfo {author} {\bibfnamefont {C.~E.}\ \bibnamefont
  {Wieman}},\ }\bibfield  {title} {\bibinfo {title} {\textrm{Atom–molecule
  coherence in a Bose–Einstein condensate}},\ }\href
  {https://doi.org/10.1038/417529a} {\bibfield  {journal} {\bibinfo  {journal}
  {Nature}\ }\textbf {\bibinfo {volume} {417}},\ \bibinfo {pages} {529–533}
  (\bibinfo {year} {2002})}\BibitemShut {NoStop}%
\bibitem [{\citenamefont {Mayle}\ \emph {et~al.}(2012)\citenamefont {Mayle},
  \citenamefont {Ruzic},\ and\ \citenamefont {Bohn}}]{PhysRevA.85.062712}%
  \BibitemOpen
  \bibfield  {author} {\bibinfo {author} {\bibfnamefont {M.}~\bibnamefont
  {Mayle}}, \bibinfo {author} {\bibfnamefont {B.~P.}\ \bibnamefont {Ruzic}},\
  and\ \bibinfo {author} {\bibfnamefont {J.~L.}\ \bibnamefont {Bohn}},\
  }\bibfield  {title} {\bibinfo {title} {Statistical aspects of ultracold
  resonant scattering},\ }\href {https://doi.org/10.1103/PhysRevA.85.062712}
  {\bibfield  {journal} {\bibinfo  {journal} {Phys. Rev. A}\ }\textbf {\bibinfo
  {volume} {85}},\ \bibinfo {pages} {062712} (\bibinfo {year}
  {2012})}\BibitemShut {NoStop}%
\bibitem [{\citenamefont {Ni}\ \emph {et~al.}(2008)\citenamefont {Ni},
  \citenamefont {Ospelkaus}, \citenamefont {de~Miranda}, \citenamefont {Peer},
  \citenamefont {Neyenhuis}, \citenamefont {Zirbel}, \citenamefont
  {Kotochigova}, \citenamefont {Julienne}, \citenamefont {Jin},\ and\
  \citenamefont {Ye}}]{KRb1}%
  \BibitemOpen
  \bibfield  {author} {\bibinfo {author} {\bibfnamefont {K.-K.}\ \bibnamefont
  {Ni}}, \bibinfo {author} {\bibfnamefont {S.}~\bibnamefont {Ospelkaus}},
  \bibinfo {author} {\bibfnamefont {M.~H.~G.}\ \bibnamefont {de~Miranda}},
  \bibinfo {author} {\bibfnamefont {A.}~\bibnamefont {Peer}}, \bibinfo {author}
  {\bibfnamefont {B.}~\bibnamefont {Neyenhuis}}, \bibinfo {author}
  {\bibfnamefont {J.~J.}\ \bibnamefont {Zirbel}}, \bibinfo {author}
  {\bibfnamefont {S.}~\bibnamefont {Kotochigova}}, \bibinfo {author}
  {\bibfnamefont {P.~S.}\ \bibnamefont {Julienne}}, \bibinfo {author}
  {\bibfnamefont {D.~S.}\ \bibnamefont {Jin}},\ and\ \bibinfo {author}
  {\bibfnamefont {J.}~\bibnamefont {Ye}},\ }\bibfield  {title} {\bibinfo
  {title} {\textrm{A High Phase-Space-Density Gas of Polar Molecules}},\ }\href
  {https://doi.org/10.1126/science.1163861} {\bibfield  {journal} {\bibinfo
  {journal} {Science}\ }\textbf {\bibinfo {volume} {322}},\ \bibinfo {pages}
  {231} (\bibinfo {year} {2008})}\BibitemShut {NoStop}%
\bibitem [{\citenamefont {Takekoshi}\ \emph {et~al.}(2014)\citenamefont
  {Takekoshi}, \citenamefont {Reichs\"ollner}, \citenamefont {Schindewolf},
  \citenamefont {Hutson}, \citenamefont {Le~Sueur}, \citenamefont {Dulieu},
  \citenamefont {Ferlaino}, \citenamefont {Grimm},\ and\ \citenamefont
  {N\"agerl}}]{GsDiMo87Rb133Cs2014Grimm}%
  \BibitemOpen
  \bibfield  {author} {\bibinfo {author} {\bibfnamefont {T.}~\bibnamefont
  {Takekoshi}}, \bibinfo {author} {\bibfnamefont {L.}~\bibnamefont
  {Reichs\"ollner}}, \bibinfo {author} {\bibfnamefont {A.}~\bibnamefont
  {Schindewolf}}, \bibinfo {author} {\bibfnamefont {J.~M.}\ \bibnamefont
  {Hutson}}, \bibinfo {author} {\bibfnamefont {C.~R.}\ \bibnamefont
  {Le~Sueur}}, \bibinfo {author} {\bibfnamefont {O.}~\bibnamefont {Dulieu}},
  \bibinfo {author} {\bibfnamefont {F.}~\bibnamefont {Ferlaino}}, \bibinfo
  {author} {\bibfnamefont {R.}~\bibnamefont {Grimm}},\ and\ \bibinfo {author}
  {\bibfnamefont {H.-C.}\ \bibnamefont {N\"agerl}},\ }\bibfield  {title}
  {\bibinfo {title} {\textrm{Ultracold Dense Samples of Dipolar RbCs Molecules
  in the Rovibrational and Hyperfine Ground State}},\ }\href
  {https://doi.org/10.1103/PhysRevLett.113.205301} {\bibfield  {journal}
  {\bibinfo  {journal} {Phys. Rev. Lett.}\ }\textbf {\bibinfo {volume} {113}},\
  \bibinfo {pages} {205301} (\bibinfo {year} {2014})}\BibitemShut {NoStop}%
\bibitem [{\citenamefont {Molony}\ \emph {et~al.}(2014)\citenamefont {Molony},
  \citenamefont {Gregory}, \citenamefont {Ji}, \citenamefont {Lu},
  \citenamefont {K\"oppinger}, \citenamefont {Le~Sueur}, \citenamefont
  {Blackley}, \citenamefont {Hutson},\ and\ \citenamefont
  {Cornish}}]{GsDiMo87Rb133Cs2014Cornish}%
  \BibitemOpen
  \bibfield  {author} {\bibinfo {author} {\bibfnamefont {P.~K.}\ \bibnamefont
  {Molony}}, \bibinfo {author} {\bibfnamefont {P.~D.}\ \bibnamefont {Gregory}},
  \bibinfo {author} {\bibfnamefont {Z.}~\bibnamefont {Ji}}, \bibinfo {author}
  {\bibfnamefont {B.}~\bibnamefont {Lu}}, \bibinfo {author} {\bibfnamefont
  {M.~P.}\ \bibnamefont {K\"oppinger}}, \bibinfo {author} {\bibfnamefont
  {C.~R.}\ \bibnamefont {Le~Sueur}}, \bibinfo {author} {\bibfnamefont {C.~L.}\
  \bibnamefont {Blackley}}, \bibinfo {author} {\bibfnamefont {J.~M.}\
  \bibnamefont {Hutson}},\ and\ \bibinfo {author} {\bibfnamefont {S.~L.}\
  \bibnamefont {Cornish}},\ }\bibfield  {title} {\bibinfo {title}
  {\textrm{Creation of Ultracold $^{87}\mathrm{Rb}^{133}\mathrm{Cs}$ Molecules
  in the Rovibrational Ground State}},\ }\href
  {https://doi.org/10.1103/PhysRevLett.113.255301} {\bibfield  {journal}
  {\bibinfo  {journal} {Phys. Rev. Lett.}\ }\textbf {\bibinfo {volume} {113}},\
  \bibinfo {pages} {255301} (\bibinfo {year} {2014})}\BibitemShut {NoStop}%
\bibitem [{\citenamefont {Park}\ \emph {et~al.}(2015)\citenamefont {Park},
  \citenamefont {Will},\ and\ \citenamefont {Zwierlein}}]{GsDiMo23Na40K2015}%
  \BibitemOpen
  \bibfield  {author} {\bibinfo {author} {\bibfnamefont {J.~W.}\ \bibnamefont
  {Park}}, \bibinfo {author} {\bibfnamefont {S.~A.}\ \bibnamefont {Will}},\
  and\ \bibinfo {author} {\bibfnamefont {M.~W.}\ \bibnamefont {Zwierlein}},\
  }\bibfield  {title} {\bibinfo {title} {\textrm{Ultracold Dipolar Gas of
  Fermionic $^{23}\mathrm{Na}^{40}\mathrm{K}$ Molecules in Their Absolute
  Ground State}},\ }\href {https://doi.org/10.1103/PhysRevLett.114.205302}
  {\bibfield  {journal} {\bibinfo  {journal} {Phys. Rev. Lett.}\ }\textbf
  {\bibinfo {volume} {114}},\ \bibinfo {pages} {205302} (\bibinfo {year}
  {2015})}\BibitemShut {NoStop}%
\bibitem [{\citenamefont {Guo}\ \emph {et~al.}(2016)\citenamefont {Guo},
  \citenamefont {Zhu}, \citenamefont {Lu}, \citenamefont {Ye}, \citenamefont
  {Wang}, \citenamefont {Vexiau}, \citenamefont {Bouloufa-Maafa}, \citenamefont
  {Qu\'em\'ener}, \citenamefont {Dulieu},\ and\ \citenamefont {Wang}}]{NaRb12}%
  \BibitemOpen
  \bibfield  {author} {\bibinfo {author} {\bibfnamefont {M.}~\bibnamefont
  {Guo}}, \bibinfo {author} {\bibfnamefont {B.}~\bibnamefont {Zhu}}, \bibinfo
  {author} {\bibfnamefont {B.}~\bibnamefont {Lu}}, \bibinfo {author}
  {\bibfnamefont {X.}~\bibnamefont {Ye}}, \bibinfo {author} {\bibfnamefont
  {F.}~\bibnamefont {Wang}}, \bibinfo {author} {\bibfnamefont {R.}~\bibnamefont
  {Vexiau}}, \bibinfo {author} {\bibfnamefont {N.}~\bibnamefont
  {Bouloufa-Maafa}}, \bibinfo {author} {\bibfnamefont {G.}~\bibnamefont
  {Qu\'em\'ener}}, \bibinfo {author} {\bibfnamefont {O.}~\bibnamefont
  {Dulieu}},\ and\ \bibinfo {author} {\bibfnamefont {D.}~\bibnamefont {Wang}},\
  }\bibfield  {title} {\bibinfo {title} {\textrm{Creation of an Ultracold Gas
  of Ground-State Dipolar $^{23}\mathrm{Na}^{87}\mathrm{Rb}$ Molecules}},\
  }\href {https://doi.org/10.1103/PhysRevLett.116.205303} {\bibfield  {journal}
  {\bibinfo  {journal} {Phys. Rev. Lett.}\ }\textbf {\bibinfo {volume} {116}},\
  \bibinfo {pages} {205303} (\bibinfo {year} {2016})}\BibitemShut {NoStop}%
\bibitem [{\citenamefont {Rvachov}\ \emph {et~al.}(2017)\citenamefont
  {Rvachov}, \citenamefont {Son}, \citenamefont {Sommer}, \citenamefont
  {Ebadi}, \citenamefont {Park}, \citenamefont {Zwierlein}, \citenamefont
  {Ketterle},\ and\ \citenamefont {Jamison}}]{GsDiMo23Na6Li2017}%
  \BibitemOpen
  \bibfield  {author} {\bibinfo {author} {\bibfnamefont {T.~M.}\ \bibnamefont
  {Rvachov}}, \bibinfo {author} {\bibfnamefont {H.}~\bibnamefont {Son}},
  \bibinfo {author} {\bibfnamefont {A.~T.}\ \bibnamefont {Sommer}}, \bibinfo
  {author} {\bibfnamefont {S.}~\bibnamefont {Ebadi}}, \bibinfo {author}
  {\bibfnamefont {J.~J.}\ \bibnamefont {Park}}, \bibinfo {author}
  {\bibfnamefont {M.~W.}\ \bibnamefont {Zwierlein}}, \bibinfo {author}
  {\bibfnamefont {W.}~\bibnamefont {Ketterle}},\ and\ \bibinfo {author}
  {\bibfnamefont {A.~O.}\ \bibnamefont {Jamison}},\ }\bibfield  {title}
  {\bibinfo {title} {\textrm{Long-Lived Ultracold Molecules with Electric and
  Magnetic Dipole Moments}},\ }\href
  {https://doi.org/10.1103/PhysRevLett.119.143001} {\bibfield  {journal}
  {\bibinfo  {journal} {Phys. Rev. Lett.}\ }\textbf {\bibinfo {volume} {119}},\
  \bibinfo {pages} {143001} (\bibinfo {year} {2017})}\BibitemShut {NoStop}%
\bibitem [{\citenamefont {Voges}\ \emph
  {et~al.}(2020{\natexlab{a}})\citenamefont {Voges}, \citenamefont {Gersema},
  \citenamefont {Meyer~zum Alten~Borgloh}, \citenamefont {Schulze},
  \citenamefont {Hartmann}, \citenamefont {Zenesini},\ and\ \citenamefont
  {Ospelkaus}}]{PhysRevLett.125.083401}%
  \BibitemOpen
  \bibfield  {author} {\bibinfo {author} {\bibfnamefont {K.~K.}\ \bibnamefont
  {Voges}}, \bibinfo {author} {\bibfnamefont {P.}~\bibnamefont {Gersema}},
  \bibinfo {author} {\bibfnamefont {M.}~\bibnamefont {Meyer~zum
  Alten~Borgloh}}, \bibinfo {author} {\bibfnamefont {T.~A.}\ \bibnamefont
  {Schulze}}, \bibinfo {author} {\bibfnamefont {T.}~\bibnamefont {Hartmann}},
  \bibinfo {author} {\bibfnamefont {A.}~\bibnamefont {Zenesini}},\ and\
  \bibinfo {author} {\bibfnamefont {S.}~\bibnamefont {Ospelkaus}},\ }\bibfield
  {title} {\bibinfo {title} {\textrm{Ultracold Gas of Bosonic
  $^{23}\mathrm{Na}^{39}\mathrm{K}$ Ground-State Molecules}},\ }\href
  {https://doi.org/10.1103/PhysRevLett.125.083401} {\bibfield  {journal}
  {\bibinfo  {journal} {Phys. Rev. Lett.}\ }\textbf {\bibinfo {volume} {125}},\
  \bibinfo {pages} {083401} (\bibinfo {year} {2020}{\natexlab{a}})}\BibitemShut
  {NoStop}%
\bibitem [{\citenamefont {Julienne}\ \emph {et~al.}(2011)\citenamefont
  {Julienne}, \citenamefont {Hanna},\ and\ \citenamefont
  {Idziaszek}}]{C1CP21270B}%
  \BibitemOpen
  \bibfield  {author} {\bibinfo {author} {\bibfnamefont {P.~S.}\ \bibnamefont
  {Julienne}}, \bibinfo {author} {\bibfnamefont {T.~M.}\ \bibnamefont
  {Hanna}},\ and\ \bibinfo {author} {\bibfnamefont {Z.}~\bibnamefont
  {Idziaszek}},\ }\bibfield  {title} {\bibinfo {title} {Universal ultracold
  collision rates for polar molecules of two alkali-metal atoms},\ }\href
  {https://doi.org/10.1039/C1CP21270B} {\bibfield  {journal} {\bibinfo
  {journal} {Phys. Chem. Chem. Phys.}\ }\textbf {\bibinfo {volume} {13}},\
  \bibinfo {pages} {19114} (\bibinfo {year} {2011})}\BibitemShut {NoStop}%
\bibitem [{\citenamefont {Ospelkaus}\ \emph {et~al.}(2010)\citenamefont
  {Ospelkaus}, \citenamefont {Ni}, \citenamefont {Wang}, \citenamefont
  {de~Miranda}, \citenamefont {Neyenhuis}, \citenamefont {Quemener},
  \citenamefont {Julienne}, \citenamefont {Bohn}, \citenamefont {Jin},\ and\
  \citenamefont {Ye}}]{Ospelkaus2010}%
  \BibitemOpen
  \bibfield  {author} {\bibinfo {author} {\bibfnamefont {S.}~\bibnamefont
  {Ospelkaus}}, \bibinfo {author} {\bibfnamefont {K.-K.}\ \bibnamefont {Ni}},
  \bibinfo {author} {\bibfnamefont {D.}~\bibnamefont {Wang}}, \bibinfo {author}
  {\bibfnamefont {M.~H.~G.}\ \bibnamefont {de~Miranda}}, \bibinfo {author}
  {\bibfnamefont {B.}~\bibnamefont {Neyenhuis}}, \bibinfo {author}
  {\bibfnamefont {G.}~\bibnamefont {Quemener}}, \bibinfo {author}
  {\bibfnamefont {P.~S.}\ \bibnamefont {Julienne}}, \bibinfo {author}
  {\bibfnamefont {J.~L.}\ \bibnamefont {Bohn}}, \bibinfo {author}
  {\bibfnamefont {D.~S.}\ \bibnamefont {Jin}},\ and\ \bibinfo {author}
  {\bibfnamefont {J.}~\bibnamefont {Ye}},\ }\bibfield  {title} {\bibinfo
  {title} {\textrm{Quantum-State Controlled Chemical Reactions of Ultracold
  Potassium-Rubidium Molecules}},\ }\href
  {https://doi.org/10.1126/science.1184121} {\bibfield  {journal} {\bibinfo
  {journal} {Science}\ }\textbf {\bibinfo {volume} {327}},\ \bibinfo {pages}
  {853} (\bibinfo {year} {2010})}\BibitemShut {NoStop}%
\bibitem [{\citenamefont {\ifmmode~\dot{Z}\else \.{Z}\fi{}uchowski}\ and\
  \citenamefont {Hutson}(2010)}]{PhysRevA.81.060703}%
  \BibitemOpen
  \bibfield  {author} {\bibinfo {author} {\bibfnamefont {P.~S.}\ \bibnamefont
  {\ifmmode~\dot{Z}\else \.{Z}\fi{}uchowski}}\ and\ \bibinfo {author}
  {\bibfnamefont {J.~M.}\ \bibnamefont {Hutson}},\ }\bibfield  {title}
  {\bibinfo {title} {Reactions of ultracold alkali-metal dimers},\ }\href
  {https://doi.org/10.1103/PhysRevA.81.060703} {\bibfield  {journal} {\bibinfo
  {journal} {Phys. Rev. A}\ }\textbf {\bibinfo {volume} {81}},\ \bibinfo
  {pages} {060703} (\bibinfo {year} {2010})}\BibitemShut {NoStop}%
\bibitem [{\citenamefont {Hu}\ \emph {et~al.}(2019)\citenamefont {Hu},
  \citenamefont {Liu}, \citenamefont {Grimes}, \citenamefont {Lin},
  \citenamefont {Gheorghe}, \citenamefont {Vexiau}, \citenamefont
  {Bouloufa-Maafa}, \citenamefont {Dulieu}, \citenamefont {Rosenband},\ and\
  \citenamefont {Ni}}]{hu2019direct}%
  \BibitemOpen
  \bibfield  {author} {\bibinfo {author} {\bibfnamefont {M.-G.}\ \bibnamefont
  {Hu}}, \bibinfo {author} {\bibfnamefont {Y.}~\bibnamefont {Liu}}, \bibinfo
  {author} {\bibfnamefont {D.~D.}\ \bibnamefont {Grimes}}, \bibinfo {author}
  {\bibfnamefont {Y.-W.}\ \bibnamefont {Lin}}, \bibinfo {author} {\bibfnamefont
  {A.~H.}\ \bibnamefont {Gheorghe}}, \bibinfo {author} {\bibfnamefont
  {R.}~\bibnamefont {Vexiau}}, \bibinfo {author} {\bibfnamefont
  {N.}~\bibnamefont {Bouloufa-Maafa}}, \bibinfo {author} {\bibfnamefont
  {O.}~\bibnamefont {Dulieu}}, \bibinfo {author} {\bibfnamefont
  {T.}~\bibnamefont {Rosenband}},\ and\ \bibinfo {author} {\bibfnamefont
  {K.-K.}\ \bibnamefont {Ni}},\ }\bibfield  {title} {\bibinfo {title}
  {\textrm{Direct observation of bimolecular reactions of ultracold KRb
  molecules}},\ }\href {https://doi.org/10.1126/science.aay9531} {\bibfield
  {journal} {\bibinfo  {journal} {Science}\ }\textbf {\bibinfo {volume}
  {366}},\ \bibinfo {pages} {1111–1115} (\bibinfo {year} {2019})}\BibitemShut
  {NoStop}%
\bibitem [{\citenamefont {Mayle}\ \emph {et~al.}(2013)\citenamefont {Mayle},
  \citenamefont {Qu\'em\'ener}, \citenamefont {Ruzic},\ and\ \citenamefont
  {Bohn}}]{PhysRevA.87.012709}%
  \BibitemOpen
  \bibfield  {author} {\bibinfo {author} {\bibfnamefont {M.}~\bibnamefont
  {Mayle}}, \bibinfo {author} {\bibfnamefont {G.}~\bibnamefont {Qu\'em\'ener}},
  \bibinfo {author} {\bibfnamefont {B.~P.}\ \bibnamefont {Ruzic}},\ and\
  \bibinfo {author} {\bibfnamefont {J.~L.}\ \bibnamefont {Bohn}},\ }\bibfield
  {title} {\bibinfo {title} {Scattering of ultracold molecules in the highly
  resonant regime},\ }\href {https://doi.org/10.1103/PhysRevA.87.012709}
  {\bibfield  {journal} {\bibinfo  {journal} {Phys. Rev. A}\ }\textbf {\bibinfo
  {volume} {87}},\ \bibinfo {pages} {012709} (\bibinfo {year}
  {2013})}\BibitemShut {NoStop}%
\bibitem [{\citenamefont {Gregory}\ \emph {et~al.}(2019)\citenamefont
  {Gregory}, \citenamefont {Frye}, \citenamefont {Blackmore}, \citenamefont
  {Bridge}, \citenamefont {Sawant}, \citenamefont {Hutson},\ and\ \citenamefont
  {Cornish}}]{Gregory_2019}%
  \BibitemOpen
  \bibfield  {author} {\bibinfo {author} {\bibfnamefont {P.~D.}\ \bibnamefont
  {Gregory}}, \bibinfo {author} {\bibfnamefont {M.~D.}\ \bibnamefont {Frye}},
  \bibinfo {author} {\bibfnamefont {J.~A.}\ \bibnamefont {Blackmore}}, \bibinfo
  {author} {\bibfnamefont {E.~M.}\ \bibnamefont {Bridge}}, \bibinfo {author}
  {\bibfnamefont {R.}~\bibnamefont {Sawant}}, \bibinfo {author} {\bibfnamefont
  {J.~M.}\ \bibnamefont {Hutson}},\ and\ \bibinfo {author} {\bibfnamefont
  {S.~L.}\ \bibnamefont {Cornish}},\ }\bibfield  {title} {\bibinfo {title}
  {\textrm{Sticky collisions of ultracold RbCs molecules}},\ }\href
  {https://doi.org/10.1038/s41467-019-11033-y} {\bibfield  {journal} {\bibinfo
  {journal} {Nature Communications}\ }\textbf {\bibinfo {volume} {10}},\
  \bibinfo {pages} {3104} (\bibinfo {year} {2019})}\BibitemShut {NoStop}%
\bibitem [{\citenamefont {Christianen}\ \emph {et~al.}(2019)\citenamefont
  {Christianen}, \citenamefont {Zwierlein}, \citenamefont {Groenenboom},\ and\
  \citenamefont {Karman}}]{PhysRevLett.123.123402}%
  \BibitemOpen
  \bibfield  {author} {\bibinfo {author} {\bibfnamefont {A.}~\bibnamefont
  {Christianen}}, \bibinfo {author} {\bibfnamefont {M.~W.}\ \bibnamefont
  {Zwierlein}}, \bibinfo {author} {\bibfnamefont {G.~C.}\ \bibnamefont
  {Groenenboom}},\ and\ \bibinfo {author} {\bibfnamefont {T.}~\bibnamefont
  {Karman}},\ }\bibfield  {title} {\bibinfo {title} {\textrm{Photoinduced
  Two-Body Loss of Ultracold Molecules}},\ }\href
  {https://doi.org/10.1103/PhysRevLett.123.123402} {\bibfield  {journal}
  {\bibinfo  {journal} {Phys. Rev. Lett.}\ }\textbf {\bibinfo {volume} {123}},\
  \bibinfo {pages} {123402} (\bibinfo {year} {2019})}\BibitemShut {NoStop}%
\bibitem [{\citenamefont {Liu}\ \emph {et~al.}(2020)\citenamefont {Liu},
  \citenamefont {Hu}, \citenamefont {Nichols}, \citenamefont {Grimes},
  \citenamefont {Karman}, \citenamefont {Guo},\ and\ \citenamefont
  {Ni}}]{liu2020steering}%
  \BibitemOpen
  \bibfield  {author} {\bibinfo {author} {\bibfnamefont {Y.}~\bibnamefont
  {Liu}}, \bibinfo {author} {\bibfnamefont {M.-G.}\ \bibnamefont {Hu}},
  \bibinfo {author} {\bibfnamefont {M.~A.}\ \bibnamefont {Nichols}}, \bibinfo
  {author} {\bibfnamefont {D.~D.}\ \bibnamefont {Grimes}}, \bibinfo {author}
  {\bibfnamefont {T.}~\bibnamefont {Karman}}, \bibinfo {author} {\bibfnamefont
  {H.}~\bibnamefont {Guo}},\ and\ \bibinfo {author} {\bibfnamefont {K.-K.}\
  \bibnamefont {Ni}},\ }\bibfield  {title} {\bibinfo {title}
  {\textrm{Photo-excitation of long-lived transient intermediates in ultracold
  reactions}},\ }\href {https://doi.org/10.1038/s41567-020-0968-8} {\bibfield
  {journal} {\bibinfo  {journal} {Nature Physics}\ }\textbf {\bibinfo {volume}
  {16}},\ \bibinfo {pages} {1132–1136} (\bibinfo {year} {2020})}\BibitemShut
  {NoStop}%
\bibitem [{\citenamefont {Gregory}\ \emph {et~al.}(2020)\citenamefont
  {Gregory}, \citenamefont {Blackmore}, \citenamefont {Bromley},\ and\
  \citenamefont {Cornish}}]{gregory2020loss}%
  \BibitemOpen
  \bibfield  {author} {\bibinfo {author} {\bibfnamefont {P.~D.}\ \bibnamefont
  {Gregory}}, \bibinfo {author} {\bibfnamefont {J.~A.}\ \bibnamefont
  {Blackmore}}, \bibinfo {author} {\bibfnamefont {S.~L.}\ \bibnamefont
  {Bromley}},\ and\ \bibinfo {author} {\bibfnamefont {S.~L.}\ \bibnamefont
  {Cornish}},\ }\bibfield  {title} {\bibinfo {title} {\textrm{Loss of Ultracold
  $^{87}\mathrm{Rb}^{133}\mathrm{Cs}$ Molecules via Optical Excitation of
  Long-Lived Two-Body Collision Complexes}},\ }\href
  {https://doi.org/10.1103/PhysRevLett.124.163402} {\bibfield  {journal}
  {\bibinfo  {journal} {Phys. Rev. Lett.}\ }\textbf {\bibinfo {volume} {124}},\
  \bibinfo {pages} {163402} (\bibinfo {year} {2020})}\BibitemShut {NoStop}%
\bibitem [{\citenamefont {Gersema}\ \emph {et~al.}(2021)\citenamefont
  {Gersema}, \citenamefont {Voges}, \citenamefont {Meyer~zum Alten~Borgloh},
  \citenamefont {Koch}, \citenamefont {Hartmann}, \citenamefont {Zenesini},
  \citenamefont {Ospelkaus}, \citenamefont {Lin}, \citenamefont {He},\ and\
  \citenamefont {Wang}}]{gersema2021probing}%
  \BibitemOpen
  \bibfield  {author} {\bibinfo {author} {\bibfnamefont {P.}~\bibnamefont
  {Gersema}}, \bibinfo {author} {\bibfnamefont {K.~K.}\ \bibnamefont {Voges}},
  \bibinfo {author} {\bibfnamefont {M.}~\bibnamefont {Meyer~zum
  Alten~Borgloh}}, \bibinfo {author} {\bibfnamefont {L.}~\bibnamefont {Koch}},
  \bibinfo {author} {\bibfnamefont {T.}~\bibnamefont {Hartmann}}, \bibinfo
  {author} {\bibfnamefont {A.}~\bibnamefont {Zenesini}}, \bibinfo {author}
  {\bibfnamefont {S.}~\bibnamefont {Ospelkaus}}, \bibinfo {author}
  {\bibfnamefont {J.}~\bibnamefont {Lin}}, \bibinfo {author} {\bibfnamefont
  {J.}~\bibnamefont {He}},\ and\ \bibinfo {author} {\bibfnamefont
  {D.}~\bibnamefont {Wang}},\ }\bibfield  {title} {\bibinfo {title}
  {\textrm{Probing Photoinduced Two-Body Loss of Ultracold Nonreactive Bosonic
  $^{23}\mathrm{Na}^{87}\mathrm{Rb}$ and $^{23}\mathrm{Na}^{39}\mathrm{K}$
  Molecules}},\ }\href {https://doi.org/10.1103/PhysRevLett.127.163401}
  {\bibfield  {journal} {\bibinfo  {journal} {Phys. Rev. Lett.}\ }\textbf
  {\bibinfo {volume} {127}},\ \bibinfo {pages} {163401} (\bibinfo {year}
  {2021})}\BibitemShut {NoStop}%
\bibitem [{\citenamefont {Bause}\ \emph {et~al.}(2021)\citenamefont {Bause},
  \citenamefont {Schindewolf}, \citenamefont {Tao}, \citenamefont {Duda},
  \citenamefont {Chen}, \citenamefont {Qu\'em\'ener}, \citenamefont {Karman},
  \citenamefont {Christianen}, \citenamefont {Bloch},\ and\ \citenamefont
  {Luo}}]{bause2021collisions}%
  \BibitemOpen
  \bibfield  {author} {\bibinfo {author} {\bibfnamefont {R.}~\bibnamefont
  {Bause}}, \bibinfo {author} {\bibfnamefont {A.}~\bibnamefont {Schindewolf}},
  \bibinfo {author} {\bibfnamefont {R.}~\bibnamefont {Tao}}, \bibinfo {author}
  {\bibfnamefont {M.}~\bibnamefont {Duda}}, \bibinfo {author} {\bibfnamefont
  {X.-Y.}\ \bibnamefont {Chen}}, \bibinfo {author} {\bibfnamefont
  {G.}~\bibnamefont {Qu\'em\'ener}}, \bibinfo {author} {\bibfnamefont
  {T.}~\bibnamefont {Karman}}, \bibinfo {author} {\bibfnamefont
  {A.}~\bibnamefont {Christianen}}, \bibinfo {author} {\bibfnamefont
  {I.}~\bibnamefont {Bloch}},\ and\ \bibinfo {author} {\bibfnamefont {X.-Y.}\
  \bibnamefont {Luo}},\ }\bibfield  {title} {\bibinfo {title} {Collisions of
  ultracold molecules in bright and dark optical dipole traps},\ }\href
  {https://doi.org/10.1103/PhysRevResearch.3.033013} {\bibfield  {journal}
  {\bibinfo  {journal} {Phys. Rev. Research}\ }\textbf {\bibinfo {volume}
  {3}},\ \bibinfo {pages} {033013} (\bibinfo {year} {2021})}\BibitemShut
  {NoStop}%
\bibitem [{\citenamefont {Nichols}\ \emph {et~al.}(2021)\citenamefont
  {Nichols}, \citenamefont {Liu}, \citenamefont {Zhu}, \citenamefont {Hu},
  \citenamefont {Liu},\ and\ \citenamefont {Ni}}]{nichols2021detection}%
  \BibitemOpen
  \bibfield  {author} {\bibinfo {author} {\bibfnamefont {M.~A.}\ \bibnamefont
  {Nichols}}, \bibinfo {author} {\bibfnamefont {Y.-X.}\ \bibnamefont {Liu}},
  \bibinfo {author} {\bibfnamefont {L.}~\bibnamefont {Zhu}}, \bibinfo {author}
  {\bibfnamefont {M.-G.}\ \bibnamefont {Hu}}, \bibinfo {author} {\bibfnamefont
  {Y.}~\bibnamefont {Liu}},\ and\ \bibinfo {author} {\bibfnamefont {K.-K.}\
  \bibnamefont {Ni}},\ }\href@noop {} {\bibinfo {title} {\textrm{Detection of
  Long-Lived Complexes in Ultracold Atom-Molecule Collisions}}} (\bibinfo
  {year} {2021}),\ \Eprint {https://arxiv.org/abs/2105.14960} {arXiv:2105.14960
  [physics.atom-ph]} \BibitemShut {NoStop}%
\bibitem [{\citenamefont {Son}\ \emph {et~al.}(2020)\citenamefont {Son},
  \citenamefont {Park}, \citenamefont {Ketterle},\ and\ \citenamefont
  {Jamison}}]{son2019observation}%
  \BibitemOpen
  \bibfield  {author} {\bibinfo {author} {\bibfnamefont {H.}~\bibnamefont
  {Son}}, \bibinfo {author} {\bibfnamefont {J.~J.}\ \bibnamefont {Park}},
  \bibinfo {author} {\bibfnamefont {W.}~\bibnamefont {Ketterle}},\ and\
  \bibinfo {author} {\bibfnamefont {A.~O.}\ \bibnamefont {Jamison}},\
  }\bibfield  {title} {\bibinfo {title} {Collisional cooling of ultracold
  molecules},\ }\href {https://doi.org/10.1038/s41586-020-2141-z} {\bibfield
  {journal} {\bibinfo  {journal} {Nature}\ }\textbf {\bibinfo {volume} {580}},\
  \bibinfo {pages} {197–200} (\bibinfo {year} {2020})}\BibitemShut {NoStop}%
\bibitem [{\citenamefont {Yang}\ \emph {et~al.}(2019)\citenamefont {Yang},
  \citenamefont {Zhang}, \citenamefont {Liu}, \citenamefont {Liu},
  \citenamefont {Nan}, \citenamefont {Zhao},\ and\ \citenamefont
  {Pan}}]{NaKYang261}%
  \BibitemOpen
  \bibfield  {author} {\bibinfo {author} {\bibfnamefont {H.}~\bibnamefont
  {Yang}}, \bibinfo {author} {\bibfnamefont {D.-C.}\ \bibnamefont {Zhang}},
  \bibinfo {author} {\bibfnamefont {L.}~\bibnamefont {Liu}}, \bibinfo {author}
  {\bibfnamefont {Y.-X.}\ \bibnamefont {Liu}}, \bibinfo {author} {\bibfnamefont
  {J.}~\bibnamefont {Nan}}, \bibinfo {author} {\bibfnamefont {B.}~\bibnamefont
  {Zhao}},\ and\ \bibinfo {author} {\bibfnamefont {J.-W.}\ \bibnamefont
  {Pan}},\ }\bibfield  {title} {\bibinfo {title} {\textrm{Observation of
  magnetically tunable Feshbach resonances in ultracold $^{23}$Na$^{40}$K
  +$^{40}$K collisions}},\ }\href {https://doi.org/10.1126/science.aau5322}
  {\bibfield  {journal} {\bibinfo  {journal} {Science}\ }\textbf {\bibinfo
  {volume} {363}},\ \bibinfo {pages} {261} (\bibinfo {year}
  {2019})}\BibitemShut {NoStop}%
\bibitem [{\citenamefont {Wang}\ \emph {et~al.}(2021)\citenamefont {Wang},
  \citenamefont {Frye}, \citenamefont {Su}, \citenamefont {Cao}, \citenamefont
  {Liu}, \citenamefont {Zhang}, \citenamefont {Yang}, \citenamefont {Hutson},
  \citenamefont {Zhao}, \citenamefont {Bai},\ and\ \citenamefont
  {Pan}}]{wang2021magnetic}%
  \BibitemOpen
  \bibfield  {author} {\bibinfo {author} {\bibfnamefont {X.-Y.}\ \bibnamefont
  {Wang}}, \bibinfo {author} {\bibfnamefont {M.~D.}\ \bibnamefont {Frye}},
  \bibinfo {author} {\bibfnamefont {Z.}~\bibnamefont {Su}}, \bibinfo {author}
  {\bibfnamefont {J.}~\bibnamefont {Cao}}, \bibinfo {author} {\bibfnamefont
  {L.}~\bibnamefont {Liu}}, \bibinfo {author} {\bibfnamefont {D.-C.}\
  \bibnamefont {Zhang}}, \bibinfo {author} {\bibfnamefont {H.}~\bibnamefont
  {Yang}}, \bibinfo {author} {\bibfnamefont {J.~M.}\ \bibnamefont {Hutson}},
  \bibinfo {author} {\bibfnamefont {B.}~\bibnamefont {Zhao}}, \bibinfo {author}
  {\bibfnamefont {C.-L.}\ \bibnamefont {Bai}},\ and\ \bibinfo {author}
  {\bibfnamefont {J.-W.}\ \bibnamefont {Pan}},\ }\href@noop {} {\bibinfo
  {title} {\textrm{Magnetic Feshbach resonances in collisions of
  $^{23}$Na$^{40}$K with $^{40}$K}}} (\bibinfo {year} {2021}),\ \Eprint
  {https://arxiv.org/abs/2103.07130} {arXiv:2103.07130 [physics.atom-ph]}
  \BibitemShut {NoStop}%
\bibitem [{Note1()}]{Note1}%
  \BibitemOpen
  \bibinfo {note} {Note, we define $2\sigma $ as the distance, where the
  density drops to $\protect \frac {1}{e^2}\cdot n(0,0,0,0)$}\BibitemShut
  {NoStop}%
\bibitem [{\citenamefont {Kraemer}(2006)}]{KraemerPhillip-Tobias2006Fiia}%
  \BibitemOpen
  \bibfield  {author} {\bibinfo {author} {\bibfnamefont {P.-T.}\ \bibnamefont
  {Kraemer}},\ }\href {http://data.onb.ac.at/rec/AC05033938} {\bibinfo {title}
  {\textrm{Few-body interactions in an ultracold gas of cesium atoms}}}
  (\bibinfo {year} {2006})\BibitemShut {NoStop}%
\bibitem [{\citenamefont {Schulze}\ \emph {et~al.}(2018)\citenamefont
  {Schulze}, \citenamefont {Hartmann}, \citenamefont {Voges}, \citenamefont
  {Gempel}, \citenamefont {Tiemann}, \citenamefont {Zenesini},\ and\
  \citenamefont {Ospelkaus}}]{SchulzeBEC2018}%
  \BibitemOpen
  \bibfield  {author} {\bibinfo {author} {\bibfnamefont {T.~A.}\ \bibnamefont
  {Schulze}}, \bibinfo {author} {\bibfnamefont {T.}~\bibnamefont {Hartmann}},
  \bibinfo {author} {\bibfnamefont {K.~K.}\ \bibnamefont {Voges}}, \bibinfo
  {author} {\bibfnamefont {M.~W.}\ \bibnamefont {Gempel}}, \bibinfo {author}
  {\bibfnamefont {E.}~\bibnamefont {Tiemann}}, \bibinfo {author} {\bibfnamefont
  {A.}~\bibnamefont {Zenesini}},\ and\ \bibinfo {author} {\bibfnamefont
  {S.}~\bibnamefont {Ospelkaus}},\ }\bibfield  {title} {\bibinfo {title}
  {Feshbach spectroscopy and dual-species \textrm{B}ose-\textrm{E}instein
  condensation of $^{23}\mathrm{Na}\text{\ensuremath{-}}^{39}\mathrm{K}$
  mixtures},\ }\href {https://doi.org/10.1103/PhysRevA.97.023623} {\bibfield
  {journal} {\bibinfo  {journal} {Phys. Rev. A}\ }\textbf {\bibinfo {volume}
  {97}},\ \bibinfo {pages} {023623} (\bibinfo {year} {2018})}\BibitemShut
  {NoStop}%
\bibitem [{\citenamefont {Voges}\ \emph
  {et~al.}(2020{\natexlab{b}})\citenamefont {Voges}, \citenamefont {Gersema},
  \citenamefont {Hartmann}, \citenamefont {Schulze}, \citenamefont {Zenesini},\
  and\ \citenamefont {Ospelkaus}}]{Voges2019Fesh}%
  \BibitemOpen
  \bibfield  {author} {\bibinfo {author} {\bibfnamefont {K.~K.}\ \bibnamefont
  {Voges}}, \bibinfo {author} {\bibfnamefont {P.}~\bibnamefont {Gersema}},
  \bibinfo {author} {\bibfnamefont {T.}~\bibnamefont {Hartmann}}, \bibinfo
  {author} {\bibfnamefont {T.~A.}\ \bibnamefont {Schulze}}, \bibinfo {author}
  {\bibfnamefont {A.}~\bibnamefont {Zenesini}},\ and\ \bibinfo {author}
  {\bibfnamefont {S.}~\bibnamefont {Ospelkaus}},\ }\bibfield  {title} {\bibinfo
  {title} {Formation of ultracold weakly bound dimers of bosonic
  ${}^{23}\mathrm{Na}^{39}\mathrm{K}$},\ }\href
  {https://doi.org/10.1103/PhysRevA.101.042704} {\bibfield  {journal} {\bibinfo
   {journal} {Phys. Rev. A}\ }\textbf {\bibinfo {volume} {101}},\ \bibinfo
  {pages} {042704} (\bibinfo {year} {2020}{\natexlab{b}})}\BibitemShut
  {NoStop}%
\bibitem [{\citenamefont {Buhmann}\ \emph {et~al.}(2008)\citenamefont
  {Buhmann}, \citenamefont {Tarbutt}, \citenamefont {Scheel},\ and\
  \citenamefont {Hinds}}]{PhysRevA.78.052901}%
  \BibitemOpen
  \bibfield  {author} {\bibinfo {author} {\bibfnamefont {S.~Y.}\ \bibnamefont
  {Buhmann}}, \bibinfo {author} {\bibfnamefont {M.~R.}\ \bibnamefont
  {Tarbutt}}, \bibinfo {author} {\bibfnamefont {S.}~\bibnamefont {Scheel}},\
  and\ \bibinfo {author} {\bibfnamefont {E.~A.}\ \bibnamefont {Hinds}},\
  }\bibfield  {title} {\bibinfo {title} {Surface-induced heating of cold polar
  molecules},\ }\href {https://doi.org/10.1103/PhysRevA.78.052901} {\bibfield
  {journal} {\bibinfo  {journal} {Phys. Rev. A}\ }\textbf {\bibinfo {volume}
  {78}},\ \bibinfo {pages} {052901} (\bibinfo {year} {2008})}\BibitemShut
  {NoStop}%
\bibitem [{\citenamefont {Bai}\ \emph {et~al.}(2020)\citenamefont {Bai},
  \citenamefont {Li}, \citenamefont {Wang}, \citenamefont {Chen}, \citenamefont
  {Si},\ and\ \citenamefont {Cong}}]{PhysRevA.101.063605}%
  \BibitemOpen
  \bibfield  {author} {\bibinfo {author} {\bibfnamefont {Y.-P.}\ \bibnamefont
  {Bai}}, \bibinfo {author} {\bibfnamefont {J.-L.}\ \bibnamefont {Li}},
  \bibinfo {author} {\bibfnamefont {G.-R.}\ \bibnamefont {Wang}}, \bibinfo
  {author} {\bibfnamefont {Z.-B.}\ \bibnamefont {Chen}}, \bibinfo {author}
  {\bibfnamefont {B.-W.}\ \bibnamefont {Si}},\ and\ \bibinfo {author}
  {\bibfnamefont {S.-L.}\ \bibnamefont {Cong}},\ }\bibfield  {title} {\bibinfo
  {title} {Simple analytical model for high-partial-wave ultracold molecular
  collisions},\ }\href {https://doi.org/10.1103/PhysRevA.101.063605} {\bibfield
   {journal} {\bibinfo  {journal} {Phys. Rev. A}\ }\textbf {\bibinfo {volume}
  {101}},\ \bibinfo {pages} {063605} (\bibinfo {year} {2020})}\BibitemShut
  {NoStop}%
\bibitem [{\citenamefont {S\"oding}\ \emph {et~al.}(1998)\citenamefont
  {S\"oding}, \citenamefont {Gu\'ery-Odelin}, \citenamefont {Desbiolles},
  \citenamefont {Ferrari},\ and\ \citenamefont {Dalibard}}]{sonding1998}%
  \BibitemOpen
  \bibfield  {author} {\bibinfo {author} {\bibfnamefont {J.}~\bibnamefont
  {S\"oding}}, \bibinfo {author} {\bibfnamefont {D.}~\bibnamefont
  {Gu\'ery-Odelin}}, \bibinfo {author} {\bibfnamefont {P.}~\bibnamefont
  {Desbiolles}}, \bibinfo {author} {\bibfnamefont {G.}~\bibnamefont
  {Ferrari}},\ and\ \bibinfo {author} {\bibfnamefont {J.}~\bibnamefont
  {Dalibard}},\ }\bibfield  {title} {\bibinfo {title} {Giant spin relaxation of
  an ultracold cesium gas},\ }\href
  {https://doi.org/10.1103/PhysRevLett.80.1869} {\bibfield  {journal} {\bibinfo
   {journal} {Phys. Rev. Lett.}\ }\textbf {\bibinfo {volume} {80}},\ \bibinfo
  {pages} {1869} (\bibinfo {year} {1998})}\BibitemShut {NoStop}%
\bibitem [{\citenamefont {Hensler}\ \emph {et~al.}(2003)\citenamefont
  {Hensler}, \citenamefont {Werner}, \citenamefont {Griesmaier}, \citenamefont
  {Schmidt}, \citenamefont {G\"{o}rlitz}, \citenamefont {Pfau}, \citenamefont
  {Giovanazzi},\ and\ \citenamefont {Rza\dot{z}ewski}}]{Hensler2003}%
  \BibitemOpen
  \bibfield  {author} {\bibinfo {author} {\bibfnamefont {S.}~\bibnamefont
  {Hensler}}, \bibinfo {author} {\bibfnamefont {U.~J.}\ \bibnamefont {Werner}},
  \bibinfo {author} {\bibfnamefont {A.}~\bibnamefont {Griesmaier}}, \bibinfo
  {author} {\bibfnamefont {P.}~\bibnamefont {Schmidt}}, \bibinfo {author}
  {\bibfnamefont {A.}~\bibnamefont {G\"{o}rlitz}}, \bibinfo {author}
  {\bibfnamefont {T.}~\bibnamefont {Pfau}}, \bibinfo {author} {\bibfnamefont
  {S.}~\bibnamefont {Giovanazzi}},\ and\ \bibinfo {author} {\bibfnamefont
  {K.~K.}\ \bibnamefont {Rza\.zewski}},\ }\bibfield  {title} {\bibinfo
  {title} {Dipolar relaxation in an ultra-cold gas of magnetically trapped
  chromium atom},\ }\href@noop {} {\bibfield  {journal} {\bibinfo  {journal}
  {Appl. Phys. B}\ }\textbf {\bibinfo {volume} {77}},\ \bibinfo {pages}
  {765–772} (\bibinfo {year} {2003})}\BibitemShut {NoStop}%
\bibitem [{Hut()}]{Hutsonprivate}%
  \BibitemOpen
  \href@noop {} {}\bibinfo {note} {J. M. Hutson and M. D. Frye, private
  communication (2021)}\BibitemShut {NoStop}%
\end{thebibliography}%
\end{document}